\documentstyle{article}
\input epsf.sty
\newcommand{\prd}{Phys.Rev.{\bf D}}
\newcommand{\prl}{Phys.Rev.Lett.}
\newcommand{\pl}{Phys.Lett.}

\title{Phase transition induced by a magnetic field}

\author{G.W.~Semenoff$^1$, I.A.~Shovkovy$^{2}$\thanks{
On leave of absence from Bogolyubov Institute for Theoretical
Physics, Kiev 252143, Ukraine},
     and L.C.R.~Wijewardhana$^2$\\
$^1$ Department of Physics and Astronomy,
\\ University of British Columbia\\
6224 Agricultural Road, Vancuver, \\
British Columbia V6T 1Z1, Canada \\
$^2$Physics Department, University of Cincinnati\\
          Cincinnati, Ohio 45221-0011, USA}

\begin{document}
\maketitle

\begin{abstract}
The magnetic catalysis of discrete chiral symmetry breaking in the 2+1
dimensional Nambu-Jona-Lasinio model is analyzed. A particular
attention is paid to a possible application of the effect in solid
state physics. The fermion contribution to the thermal conductivity as a
function of the dynamical mass (energy gap in the spectrum) is estimated
and is shown to be suppressed  when a non-zero order parameter develops.
\end{abstract}
\newpage

Chiral symmetry breaking induced by an external magnetic field has
recently been studied in a few 2+1 and 3+1 dimensional relativistic 
field theories \cite{GMS1,GMS2,Ng,Hong,Vic} (for some earlier results 
on the subject see Refs.~\cite{Kle,Klim,Schr}). In particular, a magnetic 
field enhances fermion condensation in the Nambu-Jona-Lasinio (NJL) 
model in 2+1 dimensions. The solution of the gap equation indicates 
that the critical 4-Fermi coupling required to generate a chiral 
symmetry breaking condensate with $\langle\bar\psi\psi\rangle \neq 0$ 
goes to zero as soon as the magnetic field is switched on. In fact, 
there are two branches of the solution depending on whether the 
effective 4-Fermi coupling $g$ is less than or greater than $g_c$, 
where $g_c$ is the critical coupling in the NJL model at zero magnetic 
field.  In the subcritical region $g <g_c$ the non-trivial symmetry 
breaking solution disappears as the field goes to zero. On the other 
hand, at $g > g_c$ the solution goes to the usual symmetry breaking 
solution in the NJL model without a magnetic field.

   In this letter we promote the hypothesis that this magnetic catalysis
of chiral symmetry breaking can be used to explain recent experimental
results of K.~Krishana et al \cite{Kri}.  They observed a magnetic field
induced phase transition in a high-T$_c$ superconductor (below $T_c$). The
experiment shows that the thermal conductivity as a function of
temperature and magnetic field displays a sharp break in its slope. There
are several competing arguments which attempt to explain this behavior.
In one line of reasoning, the behavior is accounted for the formation of
an energy gap at the Fermi surface. In d-wave superconductors there is no
energy gap in the spectrum of the quasi-electrons.  The Fermi surface
consists of four independent degeneracy points. The gap (as well as the
temperature of the transition) as a function of field strength follows a
square root law over a large range of fields. We present this as an
alternative to the work in \cite{th-con} which deals with some general
scaling properties of a model in a magnetic field, the work of Laughlin
\cite{BLa} which relates the phase transition to the appearance of an
additional parity and time-reversal symmetry violating superconducting
order parameter which may develop under specific conditions \cite{BLa1}
and other speculations that the behavior is a signal of a phase transition
in the lattice of magnetic flux lines. While our work was in progress, we
became aware of a paper by K.~Farakos and N.E.~Mavromatos \cite{FM} where
a similar attempt of applying the magnetic catalysis to the explanation of
the phase transition observed by Krishana et al \cite{Kri} was made. They
considered the magnetic catalysis in a gauge theory with a continuous 
symmetry, while we study the NJL model with discrete symmetry. We also
estimate the fermion contribution to the thermal conductivity as a function
of the dynamical mass, and show that it is suppressed when the order parameter
develops a nonzero expectation value.

In our hypothesis, the additional order parameter which signals the phase
transition is a neutral condensate of quasi-electron--hole pairs around
the nodes on the Fermi surface in a d-wave superconductor. The distinctive
feature is the neutrality of the condensate, and the parity as well as
time-reversal symmetry conserving character of the order parameter (same
as in \cite{FM}). The phase transition itself is of second order.

The NJL model is a natural candidate for the description of the
interactions of quasi-electrons in a d-wave superconductor. It is
well-known that, when the Fermi level is concentrated at points, the
spectrum of the quasi-electrons with energies close to the Fermi level is
linear in the momentum $E(\vec{k})\sim |\vec{k}|$ and can be described by
a Dirac Hamiltonian \cite{FM,SW,DorM}.  The effective action from which
the Dirac Hamiltonian can be derived is the Lorentz invariant Dirac action
with two species of 4-component spinors. The two species arise from the
existence of four inequivalent degeneracy points on the Fermi surface.
The Dirac action has apparent continuous chiral and flavor symmetry which
are not a symmetries of the original system which is defined on a lattice.

The interaction term is a 4-Fermi operator.  The choice of suitable
operators is dictated by the symmetries of the theory.  Though the lattice
theory does not have continuous chiral symmetry, it does have a symmetry
which corresponds to discrete chiral symmetry --- invariance under
translations by one site on the lattice.  Translation by one site
interchanges the degeneracy points of the Fermi surface and corresponds to
a discrete chiral transformation in the continuum. For this reason it is
natural to choose a 4-Fermi interaction which breaks the continuous chiral
symmetry to its discrete $Z_2$ subgroup.

It is also natural to choose a 4-Fermi operator which is invariant under
parity, time reversal and charge conjugation invariance. Parity and time
reversal are symmetries of the theory in the absence of an external
magnetic field and charge-conjugation invariance is equivalent to the
approximate particle-hole symmetry of the low energy excitations.

Even with these restrictions, there are both Lorentz invariant and
non-invariant 4-Fermi operators.  The rationale for choosing a Lorentz
invariant operator is that, it is a combination of Lorentz invariance and
chiral symmetry which forbids an energy gap in the spectrum of the
continuum model.  If we chose a non Lorentz invariant 4-Fermi operator,
radiative corrections would gap the quasi-electron spectrum even without
an external magnetic field or spontaneous symmetry breaking.  A candidate
for generating 4-Fermi interactions of the kind that we choose, $-g
\left(\bar\psi\psi\right)^2$ is the effective interaction mediated by
optical phonons.

Consider the Nambu-Jona-Lasinio model in 2+1 dimensions \cite{SW,NJL}. In
2+1 dimensions the minimal representation of the Dirac algebra is with $2
\times 2$ matrices and the minimal Dirac spinor has two components.
However, for two component fermions, the mass operator $\bar\psi\psi$ is a
pseudo-scalar and massive fermions violate parity. If there are two
species of fermions, the mass operator $\bar\psi_1\psi_1-\bar\psi_2\psi_2$
is a scalar if, besides the spacetime transformation, the fermion species
are interchanged under parity. The other combination,
$\bar\psi_1\psi_1+\bar\psi_2\psi_2$ is a pseudoscalar.  It is convenient
to describe the two species of fermions using a reducible $4\times 4$
representation if the Dirac algebra \cite{ABKW}. The $\gamma$-matrices
are:
\begin{equation}
\gamma^0=
         \left(\begin{array}{ll}
         \sigma_3 &  0       \\
          0       & -\sigma_3
         \end{array}\right), \quad
\gamma^1=
         \left(\begin{array}{ll}
         i\sigma_1 &  0       \\
         0       & -i\sigma_1
         \end{array}\right), \quad
\gamma^2=
         \left(\begin{array}{ll}
         i\sigma_2 &  0       \\
         0       & -i\sigma_2
         \end{array}\right),
\label{gamma-m}
\end{equation}
where $\sigma_i$ are the Pauli matrices.

Thus, the Lagrangian density of our model is given by
\begin{equation}
{\cal L} =\frac{1}{2}
\left[\bar{\Psi}, i\gamma^\mu D_\mu\Psi\right]
+ \frac{g}{2} (\bar{\Psi}\Psi)^2,
\label{lagNJL}
\end{equation}
where the covariant derivative includes only the vector
potential for the external magnetic field:
\begin{equation}
D_\mu=\partial_\mu-ie A^{ext}_\mu,\quad
A^{ext}_\mu=\left(0,-\frac{B}{2}x_2,\frac{B}{2}x_1,0\right).
\label{dmu1}
\end{equation}
To apply the $1/N$ perturbative expansion we assume that fermions have
an additional flavor index $i=1,2,\dots,N$.  In the d-wave superconductor
N=2.

The standard mass term, $m\bar{\Psi}\Psi$, as is easy to check, is not
invariant under the discrete chiral symmetry
\begin{equation}
\Psi\to\gamma_5\Psi, \quad
\bar{\Psi} \to - \bar{\Psi} \gamma_5.
\label{sym}
\end{equation}
So, it is forbidden in the Lagrangian density. However, due to the
magnetic catalysis \cite{GMS1} the discrete symmetry is  broken
dynamically  and as a result, fermions get a dynamical mass.
Let us stress that symmetry breaking appears already at an arbitrary
small 4-fermion coupling constant.

The latter fact can be understood by the following argument.
Consider the Dirac Hamiltonian in a background magnetic field
\begin{equation}
H=\left( \matrix{h & 0\cr 0& h }\right)
\end{equation}
where
\begin{equation}
h=i\sigma^2D_1-i\sigma^1D_2.
\end{equation}

In a background field with magnetic flux $\Phi=\frac{1}{2\pi}\int Bd^2x$
an index theorem (an open space version of the Atiyah-Singer Index
theorem) and a vanishing theorem imply that there are $[|\Phi|]-1$
solutions of the equation $h\phi=0$ where $[|\Phi|]$ is the largest integer
less than $|\Phi|$, and their chirality is given by $\sigma^3\phi={\rm
sign}(\Phi)([|\Phi|]-1)$. Existence of these zero modes of the Hamiltonian
implies degeneracy of the second quantized ground state - the degenerate
states correspond to all of the ways that the zero modes could be
populated by quasi-electrons. Overall charge neutrality implies that half
of the zero modes should be populated and half of them should be empty.
There are chirally symmetric populations of the zero modes where half of
the modes with four component eigenfunctions of $H$, $\Psi_+=
\left(\matrix{ \phi\cr 0\cr}\right)$ and half of the modes with
wavefunctions $\Psi_-=\left( \matrix{ 0\cr \phi\cr}\right)$ are occupied.
The chiral condensate is $ \langle\Psi^{\dagger}\gamma^0\Psi \rangle$, 
and the average condensate 
$\langle \int d^2x \Psi^{\dagger}\gamma^0\Psi \rangle $
vanishes.  There are also chirally asymmetric populations of the zero
modes, for example, when all of the positive helicity modes $\Psi_+$ are
filled and the negative helicity modes $\Psi_-$ are empty.  They are
degenerate with the chirally symmetric ones.

The degeneracy is resolved by adding the interaction terms in the
Hamiltonian and diagonalizing the matrix of first order corrections.  In
this case, the perturbation given by the 4-Fermi operator $-G/2\left(
\bar\Psi\Psi\right)^2$ resolves the degeneracy by lowering the energy of
the states with maximally chirally asymmetric populations. It is also easy
to see that a similar argument would not work when there is no magnetic
field since in that case there are no zero modes and the density of states
at the Fermi surface vanishes like $|E-E_F|$.  The excited states which
would have to be populated in order to have a non-vanishing condensate
have positive energy and therefore there is a finite critical coupling
which $g$ must achieve before the 4-Fermi can lower the energy of the
chirally asymmetric state below that of the ground state.

Conventionally, the symmetry breaking is found using the effective
potential in the large $N$ expansion. The analysis of the model with the
Lagrangian density in (\ref{lagNJL}) repeats, up to minor differences
appearing due to the absence of Nambu-Goldstone bosons, the analysis in
\cite{GMS1}. We shall skip all of the intermediate calculations and go
directly to the gap equation
\begin{equation}
\mu l =\frac{1}{lm_{dyn}} +\sqrt{2}
\zeta(\frac{1}{2},1+\frac{m_{dyn}^2l^2}{2})+O(1/\Lambda).
\label{eq:gap}
\end{equation}
Here we introduced the mass scale parameter $\mu\equiv
2\Lambda(g_c-g)/gg_c$ (by definition, $g=NG\Lambda/\pi$ and
$g_c=\sqrt{\pi}$). In addition, the magnetic length $l \equiv
1/\sqrt{|eB|} $ is used instead of the magnetic field strength $B$. From
now on we assume that the value of cutoff $\Lambda$ is much bigger than
all other parameters in the model, so that further we can neglect all
corrections of order $1/\Lambda$ or higher.

The gap equation (\ref{eq:gap}) is written for the ``subcritical" value of
coupling constant, $g\leq g_c$ (where $g_c$ is the critical value of
coupling in the model without the magnetic field). If we try to go into
the supercritical region the parameter $\mu$ changes the sign. In this
paper, however, we are not going to study the case when $g>g_c$. As is
clear, by fixing the dimensionful parameter $\mu$, one already defines the
NJL model. As soon as the $\mu$ is fixed, we can talk about strong or weak
external magnetic fields applied to the system. The quantitative
characteristics of that is the value of dimensionless parameter $\mu l$.
Indeed, the vanishing field will correspond to infinitely large value of
$\mu l$. While increasing the field, the parameter $\mu l$ gets smaller,
and eventually goes to zero when field becomes infinitely strong. Of
course, different values of $\mu l$ can also be interpreted from another
viewpoint: small parameter $\mu l$ corresponds to near-critical value of
coupling constant, while large value of $\mu l$ corresponds to weak
coupling.

For our analysis below, the first approach seems to be more appropriate
since we are interested in the dependence of dynamical mass on the
magnetic field strength without changing the ``parent" NJL model. It would
also allow us to make the analysis almost independent of a specific value
of coupling constant.

So, let us consider, two limiting cases of field strength: (i) vanishingly
weak field (or rather $\sqrt{|eB|}\ll \mu$), and (ii) infinitely large
field (or just $\sqrt{|eB|}\gg \mu$).

To get the gap equation in the case of weak field, one has to notice that
large positive value in the right hand side of (\ref{eq:gap}) is obtained
for small values of $m_{dyn}l$ when the first term dominates over the zeta
function term. Thus, the approximate gap equation reads
\begin{equation}
\mu l \simeq  \frac{1}{lm_{dyn}},
\end{equation}
so that the solution for the dynamical mass is proportional to the
magnetic field strength,
\begin{equation}
m_{dyn}=\frac{|eB|}{\mu}.
\end{equation}
Here we would like to note that a simple numerical analysis of the
gap equation (\ref{eq:gap}) shows that the solution is indeed close
to this asymptotic behavior up to a few percent but only in the region 
where the ratio $\sqrt{|eB|}/\mu$ is less than around 0.005. If the 
ratio gets bigger than that deviations are more prominent.

The case of strong magnetic field is slightly more complicated. The first
approximation for the gap equation in this case is obtained by
substituting zero in the left hand side of the gap equation
(\ref{eq:gap}):
\begin{equation}
0=\frac{1}{lm_{dyn}} +\sqrt{2}
\zeta(\frac{1}{2},1+\frac{m_{dyn}^2l^2}{2}).
\label{gap-2}
\end{equation}
This equation can easily be solved numerically, leading to the result
$m_{dyn}l=c_0\approx 0.446$. And the solution for the dynamical mass is
proportional to the square root of the field,
\begin{equation}
m_{dyn}\simeq c_0\sqrt{|eB|}+c_1\mu,
\label{asy-so}
\end{equation}
where the second term (with $c_1=-0.173$) is the next to leading term
in Taylor expansion around $\mu l=0$. It appears when the magnetic
field strength is large but finite. Even though the solution in
(\ref{asy-so}) was obtained under the assumption $\sqrt{|eB|}\gg \mu$,
it should be rather reliable even for values of $\sqrt{|eB|}$ of order
$\mu$. In fact, the numerical solution of the gap equation, given in
Figure~1, confirms our statement. More than that, as a simple analysis
shows, the numerical solution can be well approximated by a the square
root dependence in the whole region of magnetic fields except for a
small region of weak fields.

Thus, summarizing, the dependence of the dynamical mass on the external
magnetic field is linear for very small fields (small in comparison with
the inherent scale $\mu$), then the dependence gradually approaches the
asymptote in Eq.~(\ref{asy-so}) when the square root of field gets bigger
than $\mu$.  The graphical solution is shown in Figure~1.

In view of application the result to the phase transition mentioned in the
introduction, we have to require that the actual value of the parameter
$\mu$ is small enough in comparison with the square root of the magnetic
field while given in the appropriate units. So, that the small linear
region of the dynamical mass dependence is not experimentally seen. This
assumption is not unnatural at all if we recall that the asymptotic behavior 
in (\ref{asy-so}) has a rather wide range of validity, while the linear
dependence appears only for really very weak fields.

The gap equation at a finite temperature can also be written right away
\cite{GMS1}. However, in that case it becomes really complicated and we
are not going to write it down here. The qualitative picture though is
clear enough even without solving the equation. In particular, we know
that the magnetic field helps symmetry breaking, while temperature works
against it. As a result, we have a competition of those two influences. If
we keep the value of the field being constant, the critical temperature at
which the order parameter (gap in the energy spectrum) disappears is, up
to an unimportant numerical factor, proportional to the value of the
dynamical mass at zero temperature $m_{dyn}(|eB|)$ \cite{GMS1,Ng,GS}. On
the other hand, considering the mass as a function of the field, we can
invert the equation for the critical temperature, and obtain the critical
value of the magnetic field at a given temperature. So, looking for the
dependence of the dynamical mass on the field strength at a finite
temperature, we will find that the gap stays zero until the value of the
field is less than critical, and after field becomes stronger than
critical, a non-zero gap develops and grows with increasing the field.
Later we will make use of this qualitative picture.

We would like to mention that one may be tempted and try to use the
magnetic catalysis of 2+1 dimensional quantum electrodynamics instead of
the NJL model. It is indeed a very attractive idea since, in a rather
physical limit of a weak magnetic field (and apparently in range of
validity of the large $N$ expansion), the dynamical mass as a function of
the magnetic field follows the square root law dependence \cite{Shpa}:
\begin{equation}
m_{dyn}\simeq\frac{\sqrt{|eB|}}{4\pi\nu_0N}\ln(4\pi\nu_0N),
\label{weak}
\end{equation}
where $\nu_0\approx 0.14$. Note that this dependence is quite natural from
the following point of view. As is well known, in 2+1 dimensional QED it
is $1/N$ rather than $e^2$ that plays the role of the coupling constant
\cite{ABKW,ANW}. The value of $e^2$ is more like an ultraviolet cutoff in
the model. Therefore, one effectively is left with a single infrared
dimensionful parameter, namely $\sqrt{|eB|}$, and the dynamical mass 
should be proportional to this only one dimensionful parameter.

While considering quantum electrodynamics one should remember, however,
that breaking the continuous symmetry is forbidden in 2+1 dimensions at
any finite temperature. In order to avoid this difficulty one should apply
additional assumptions, like those in \cite{FM} where the realization of
the BKT phase is conjectured.

Another thing that can obscure the effect of magnetic catalysis in 2+1
dimensional QED is the parity violating mass term that seems to be
generated due to the induced Chern-Simons photon mass in a magnetic field
\cite{Hos}. Even if the effect of magnetic catalysis does not disappear,
the quantitative analysis may somewhat change.

It is possible to give a simple estimate of the thermal conductivity
dependence on the magnetic field in the model at hand.

To obtain the expression for the thermal conductivity we apply the
familiar method of linear response \cite{Kubo}. There are, however, a few
things that we need to mention. First, we work with zero chemical
potential, so that the general expression for the conductivity is given in
terms of the energy-current correlation function alone. And second, to get
a finite thermal conductivity we will need to modify the free fermion
Green function in the magnetic field by introducing an effective lifetime
of quasiparticles. The latter can be interpreted as a result of fermion
scattering at impurities of a superconductor, which is always present in
real samples. Such an approximation is presumably good enough for our
purposes, since the only thing we are interested in is a change in the
value of the thermal conductivity that results from the magnetic catalysis
of symmetry breaking and generation of the energy gap.

As we mentioned above, the expression for the thermal conductivity can
be given in terms of the energy-current (momentum) correlation
function \cite{Kubo}
\begin{equation}
\kappa^{el}_{ij}(\omega)=\frac{1}{TV}\int\limits_{0}^{\infty}dt
\int\limits_{0}^{\beta} d\lambda Tr\left\{ \rho_{0}
P^{i}(0)P^{j}(t+i\lambda)\right\}e^{-i\omega t},
\label{kap1}
\end{equation}
where $V$ is the volume of the system,
\begin{eqnarray}
P^{i}(0)&=&\frac{i}{2}\int d^2 x\left(
\bar{\psi} \gamma^{0} \partial^{i}\psi
-\partial^{i}\bar{\psi} \gamma^{0} \psi \right),  \label{mom}\\
P^{j}(t)&=&e^{iHt}P^{j}(0)e^{-iHt}
\end{eqnarray}
and $\rho_0$ is the equilibrium density matrix:
\begin{equation}
\rho_0=\frac{1}{Z}e^{-\beta H}, \qquad Z=Tr\left(e^{-\beta H}\right).
\label{den-mat}
\end{equation}
Notice here that the action is invariant under translations even in the
presence of a constant external magnetic field, and that the definition
of the momentum coincides with that for the theory without an external
field. Here we do not include the contribution of the gauge fields
to the momentum since we are interested in the fermion component of
the thermal conductivity.

By using the definition in Eq.~(\ref{kap1}), one can obtain the following
expression for the static thermal conductivity of an isotropic system
\cite{Lang,AT,AG}
\begin{equation}
\kappa^{el}=-\frac{1}{TV}Im\int\limits_{0}^{\infty}tdt
~Tr \left\{ \rho_{0}
P^{i}(t)P^{i}(0)\right\}.
\label{kap2}
\end{equation}
Let us introduce the following thermal Green function:
\begin{eqnarray}
G(\tau)&=&\frac{1}{V}Tr\left(\rho_{0}e^{\tau H}
 P^{i}e^{-\tau H} P^{i}\right), \label{gr-fun}\\
G(i\nu_m)&=&\int\limits_{0}^{\beta}G(\tau)e^{i\nu_m \tau} d\tau,
\end{eqnarray}
where $\nu_m=2\pi T m$.
It is this Green function $G(\nu_m)$ that can perturbatively be
calculated in a theory. The simplest Feynman diagram gives
\begin{equation}
G(p=0,i\nu_m)=iT\sum_{n=-\infty}^{+\infty}
\int \frac{d^2 k}{(2\pi)^2} k^2
\mbox{tr}\left(\gamma^{0}S(i\omega_{n},k)
\gamma^{0}S(i\omega_{n}+i\nu_m,k)\right),
\label{loop}
\end{equation}
It turns out that the thermal conductivity can be given by the
discontinuity of this Green function (see \cite{Lang,AT,AG} for details)
\begin{equation}
\kappa^{el}=\frac{1}{4T}\lim_{\omega\to 0}\frac{1}{\omega}
\left(G(p=0,i\nu_m=\omega+i0^{+})-G(p=0,i\nu_m=\omega-i0^{+})\right).
\label{kap3}
\end{equation}
To calculate the Green function in Eq.~(\ref{loop}), it is convenient to
use the following spectral representation for the fermion thermal Green
function \cite{AT,AG}
\begin{equation}
S(i\omega_{n},k)=\int\limits_{-\infty}^{+\infty}\frac{d\omega}{2\pi}
\frac{a(\omega,k)}{i\omega_{n}-\omega},
\label{spect}
\end{equation}
where $a(\omega,k)=2Im S(i\omega_{n}=\omega-i0^{+},k)$.
After substituting this  spectral representation into
Eq.~(\ref{loop}), one can easily perform the sum over the Matsubara
frequencies. Then the result reads
\begin{eqnarray}
G(p=0,i\nu_m)&=&-\frac{i}{2}
\int \frac{d^2 k}{(2\pi)^2}\int \frac{d\omega_{1}}{2\pi}
\int \frac{d\omega_{2}}{2\pi}
\frac{k^2\sinh\left(\frac{\omega_{1}-\omega_{2}}{2T}\right)}
{ \cosh\left(\frac{\omega_{1}}{2T}\right)
\cosh\left(\frac{\omega_{2}}{2T}\right)}
\nonumber\\
&\times&\frac{\mbox{tr}\left(\gamma^{0}a(\omega_{1},k)
\gamma^{0}a(\omega_{2},k)\right)}
{\left(\omega_{1}-\omega_{2}-i\nu_m\right) }.
\label{loop1}
\end{eqnarray}
As we see, this representation is indeed very convenient for extracting the
discontinuity that is needed for calculation of the thermal conductivity
[see Eq.~(\ref{kap3})]
\begin{equation}
\kappa^{el}=\frac{1}{16T^2}
\int \frac{d^2 k}{(2\pi)^2}\int \frac{d\omega}{2\pi}
\frac{k^2}{ \cosh^{2}\left(\frac{\omega}{2T}\right) }
\mbox{tr} \left(\gamma^{0}a(\omega,k)
\gamma^{0}a(\omega,k)\right).
\label{kap4}
\end{equation}
First, let us calculate the thermal conductivity in the lowest Landau
level approximation without taking into account any interactions in
the theory. The corresponding fermion Green function is given by
\begin{equation}
S(i\omega_{n},k)=\exp\left(-\frac{k^2}{|eB|}\right)
\frac{i\omega_{n}\gamma^{0}+m}{(i\omega_{n})^2-m^2}
\left(1-i\gamma^{1}\gamma^{2}\right),
\end{equation}
so that the spectral density reads
\begin{equation}
a(\omega,k)=2\pi \mbox{sign}(\omega)
\exp\left(-\frac{k^2}{|eB|}\right)
\delta\left(\omega^2-m^2\right)
\left[\omega\gamma^{0}+m\right]
\left(1-i\gamma^{1}\gamma^{2}\right).
\label{sp}
\end{equation}
Making use of this explicit expression for the spectral density,
one obtains  the thermal conductivity right away
\begin{equation}
\kappa^{el}=\frac{|eB|^2}{16 T^2 }
\mbox{sech}^2\left(\frac{m}{2T}\right)\delta(0).
\label{kappa}
\end{equation}
This result looks meaningless because of the  $\delta$-function in the
right hand side which is infinite. Note, however,  that this infinity
appears  due to the $\delta$-like peak of the spectral density
$a(k,\omega)$ in Eq.~(\ref{sp}). The latter is a direct consequence
of our dealing with a free theory. In practice, the spectral density
of any physical interacting model is always spread over a finite region
of energies.

Instead of actual consideration of the interaction effects on the spectral
density in our model, we just modify the latter by applying simple
phenomenological arguments. In particular, we introduce the width of
quasiparticle states (or in different words, the inverse quantity of the
lifetime), $\Gamma\equiv 1/\tau\ll m$, according to the following rule
\begin{eqnarray}
\delta\left(\omega^2-m^2\right)&=&\frac{1}{2m}\left(
\delta\left(\omega+m\right)+\delta\left(\omega-m\right)
\right) \nonumber \\
&\to& \frac{1}{2m\pi}\left(
\frac{\Gamma}{\left(\omega+m\right)^2+\Gamma^2}+
\frac{\Gamma}{\left(\omega-m\right)^2+\Gamma^2}
\right) .
\end{eqnarray}
At this point we may not specify the nature of the quasiparticle lifetime,
but use it as a phenomenological parameter. However, we note that such
broadening of the spectral density can be accounted for the scattering of
fermions off impurities which always present in real systems \cite{AT,AG}.

Now we are in a position to estimate the thermal conductivity in our
modified theory. In order to preserve the validity of our symmetry
breaking analysis due to the magnetic catalysis, we assume that the width
$\Gamma$ is considerably smaller than the dynamical mass, $\Gamma\ll m$.
Then substitution of the spectral density into the expression for the
conductivity (\ref{kap4}) leads to the final result
\begin{equation}
\kappa^{el}\simeq \frac{|eB|^2}{32\pi T^2 \Gamma}
\mbox{sech}^2\left(\frac{m}{2T}\right).
\label{kappa1}
\end{equation}
Let us apply this simplified result to the study of thermal conductivity
dependence on the magnetic field at a fixed temperature. Qualitatively, we
get the following picture. While the magnetic field is weaker than the
critical value at a given temperature, there is no energy gap in the
spectrum of the model. So, we have to substitute $m(T,|eB|<|eB_c|)=0$ in
Eq.~(\ref{kappa}). As a result, we observe that the thermal conductivity
increases with the magnetic field:
\begin{equation}
\kappa_{el}=\frac{|eB|^2}{32\pi T^2 \Gamma}.
\label{B<B_c}
\end{equation}
This seems to be just the opposite what the experiment sees \cite{Kri}.
However, we believe that this increase might be the artifact of
inappropriate use of the approximation (see below) as well as a result of
absence of the Fermi surface in our model and should not be taken as a
real prediction.

On the other hand, after the magnetic field gets bigger than the critical
value, $|eB|>|eB_c|$, the energy gap $m(T,|eB|>|eB_c|)\neq 0$ develops and
grows with increasing the field strength. So, that gradually, the
dependence of the thermal conductivity becomes exponentially damped by the
energy gap:
\begin{equation}
\kappa_{el}=\frac{|eB|^2}{8\pi T^2 \Gamma}.
\exp\left(-\frac{m(T,|eB|)}{T}\right).
\label{B>B_c}
\end{equation}
We expect that this damping of the conductivity is a rather general
feature which should be present as soon as an energy gap is developed in
the model. We remind that something similar also happens with the thermal
conductivity when an ordinary superconducting order parameter develops
\cite{AT}.

In support of the estimates made above, let us comment on the lowest
Landau level approximation that was used in the derivation of
Eq.~(\ref{kappa}). Obviously, this approximation could be reliable only if
we stay in the region where the dynamical mass is much smaller than the
scale of the magnetic field. Let us remind that this condition was indeed
satisfied in the case of the magnetic catalysis in QED at zero
temperature. While considering finite temperatures, we have to require
also that the temperature is much less than the square root of the field
strength. This latter requirement gives us one more argument in support of
reliability of result for the thermal conductivity (\ref{B>B_c}) in the
case of supercritical fields, while adds more distrust in the analysis for
subcritical magnetic fields.

Concerning the NJL model, we mention that the ratio $m/\sqrt{|eB|}$, as is
seen from Eq.~(\ref{asy-so}), approaches $0.446$ in the strong field
limit. It is this region that reproduces the square root dependence of the
thermal conductivity as a function of the magnetic field. Consider,
however, that approximately the same dependence law remains valid down to
the values of the ratio $m/\sqrt{|eB|}$ as small as $0.1$. Therefore, even
though the lowest Landau level approximation may not be perfect in case of
the NJL model, it should be reliable enough for the thermal conductivity
estimate and for the conclusion about the damping of the latter in the
supercritical region.

The study of the 2+1 dimensional relativistic NJL model, modeling the
dynamics of quasiparticles close to the nodes of the Fermi surface in
d-wave superconductors, reveals the dependence for the mass as a function
of the magnetic field very close to square root law for rather general
assumptions. The appearance of the energy gap is viewed here as a result
of breaking a discrete symmetry due to neutral condensate of
fermion-antifermion pairs. In condense matter language such a situation
could be interpreted as the condensation of electron-hole-like
quasiparticle pairs. The results of our paper give an alternative
explanation to the recently observed \cite{Kri} phase transition in
cuprites below $T_c$.

\section{Acknowledgments}

We would like to thank V.~Gusynin and V.~Miransky for discussions and
pointing our attention to Ref.~\cite{FM}. This work was supported by the
Natural Sciences and Engineering Research Council of Canada and in part by
the U.S. Department of Energy Grant \#DE-FG02-84ER40153.

\begin{figure}
\epsfbox{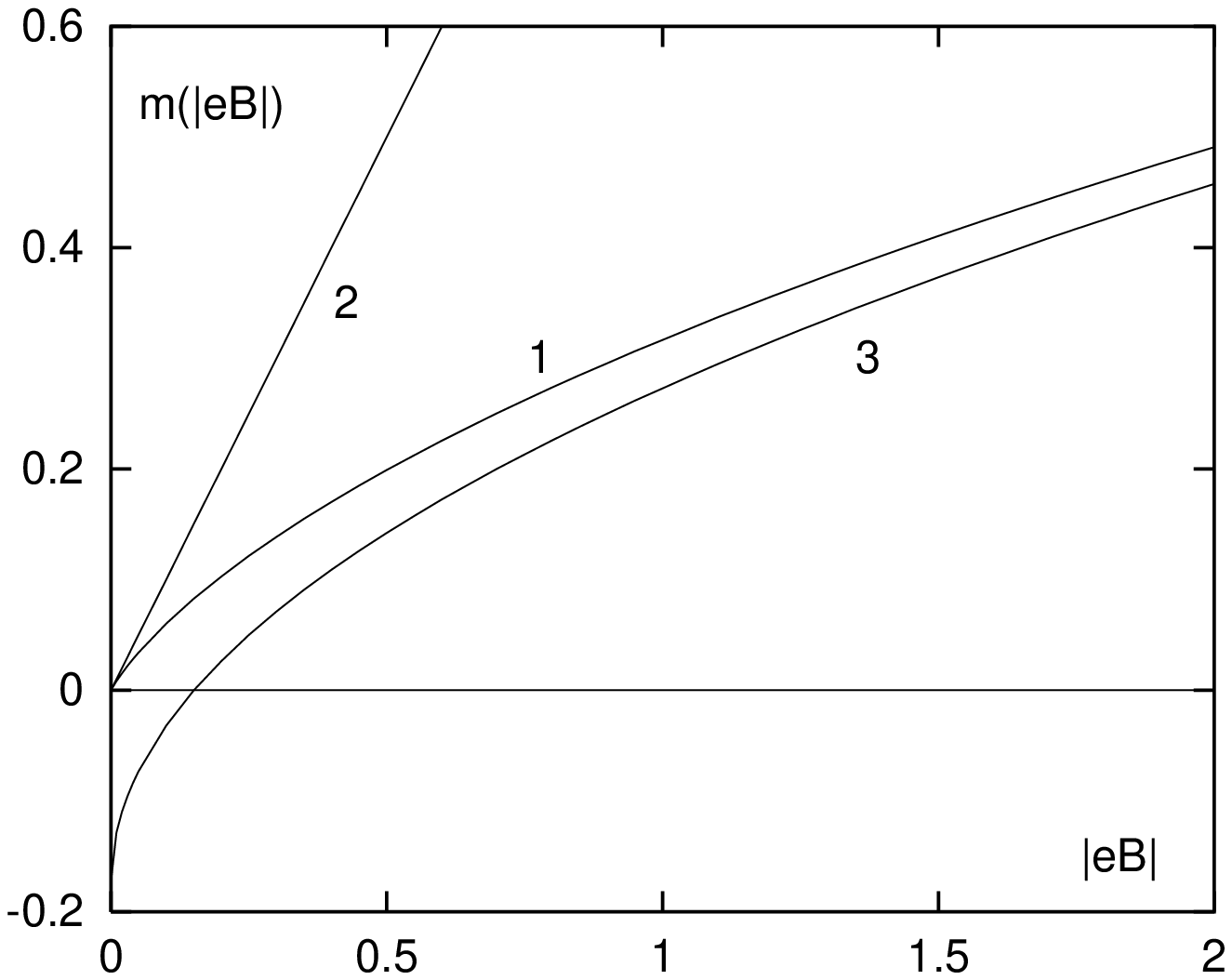}
\caption{Curve 1 is the numerical solution to the gap equation in the
NJL model. Curves 2 and 3 are the weak field and strong field
asymptotes of the exact solution, respectively. All quantities
are measured in units of $\mu$.}
\label{Figure1}
\end{figure}

\end{document}